\documentclass[aps,prd,twocolumn,groupedaddress]{revtex4}
\usepackage{graphicx}
\usepackage{dcolumn}
\usepackage{bm}
\usepackage{amsmath}
\usepackage{amssymb}
\usepackage{color}
\usepackage{slashed}
\usepackage{soul}
\usepackage{tikz}

\begin{document}

\title{Hidden momentum of composite bodies and magnetic dipoles}

\author{Bruno Klajn}\email{bklajn@zsem.hr}
\affiliation{Zagreb School of Economics and Management, Filipa Vukasovi\' ca 1, 10000 Zagreb, Croatia.}

\author{Hrvoje Nikoli\'c}\email{hnikolic@irb.hr}
\affiliation{Theoretical Physics Division, Rudjer Bo\v{s}kovi\'{c} Institute, P.O.B. 180, HR-10002 Zagreb, Croatia.}

\begin{abstract}
Hidden momentum is a puzzling phenomenon associated with magnetic dipoles and other extended relativistic systems.
We point out that the origin of hidden momentum lies in the effective change of individual particle masses
of a composite body,
during which the total momentum of the system is not equal to the momentum of the center of mass.
Defining the hidden momentum as the difference between the total momentum and the center-of-mass momentum,
we explain in detail how hidden momentum arises in certain simple non-relativistic systems,
in typical relativistic systems due to velocity-dependent ``relativistic mass'', and in 
magnetic dipoles as a special case of relativistic systems.
\end{abstract}

\maketitle

\section{Introduction} 

For a motivation, let us start with a puzzling question that can be asked at a physicists party:

- Q: A truck is at rest. How to give the truck a momentum while keeping it at rest? 

- A: Pour some earth in the truck at its rear end.
Considering the earth on the truck to be a part of the truck, the extra mass translates the center of mass  
towards the rear end. During the pouring, therefore, the center of mass has velocity 
without moving the track. The mass times velocity of the center of mass is the momentum of the center of mass,
which gives the truck a momentum while keeping it at rest.

But note that, in this example, no physical object is moving, 
the translation of the center of mass does not correspond to a motion of any 
physical object. In fact, as we shall see, this translation of the center of mass can even be faster than light.  

This example illustrates the basic idea of hidden momentum in the simplest possible terms.
But in the literature \cite{hnizdo1992, internal, babson}, hidden momentum has been introduced in much more complicated ways, as an effect 
related to magnetic dipoles and special relativity. The purpose of this paper 
is to demystify the 
hidden momentum by explaining its true origin and physical meaning.

Historically, the subtle interplay between electromagnetic fields and magnetized or charged matter was found to give rise to momentum that eludes purely mechanical consideration. Early analyses demonstrated the existence of this ``hidden momentum'' in systems involving magnets and electric currents, revealing that careful attention to energy and momentum flow is required to maintain consistency with Maxwell’s equations and conservation laws \cite{ShockleyJames1967, ColemanVanVleck1968, HausPenfield1968}. After the notion of hidden momentum became widely appreciated in the physics community, it became a unavoidable ingredient in explaining the interactions of magnetic dipoles, both in classical \cite{boyer, hnizdo1992, vaidman, internal} and quantum settings \cite{AharonovCasher1984}.

Standard texts on electrodynamics now discuss hidden momentum explicitly, illustrating its importance as an integral part of the field’s momentum and energy balance \cite{Griffiths4th}. Although the concept of hidden momentum is still almost exclusively discussed as related to electromagnetism \cite{resource, babson}, a completely general definition was recently given \cite{vanzella}. with applications far beyond electromagnetism. A plethora of interesting examples considering hidden momentum is also available online \cite{McDonaldWeb}.

In recent times, the concept of hidden momentum was used to resolve the so-called Mansuripur ``paradox'' \cite{Mansuripur2012, GriffithsHnizdo2013, Vanzella2013}, as well as to provide a better understanding of the long-lasting Abraham-Minkowski controversy \cite{Barnett2010, Barnett2013}.

The present paper is organized as follows. In the second section we investigate a system in Newtonian mechanics, 
akin to the example from the beginning of the paper, and find that, in general, mechanical systems can carry the momentum 
which is not related to the overt motion of the system. In the third section we 
explain how a similar effect appears naturally in all relativistic systems, 
and focus on systems of magnetic (and electric) dipoles in fourth section. 
We conclude our findings in the fifth section and provide details of some calculations in Appendix.

\section{Hidden momentum in Newtonian mechanics}

Consider a system of $N$ bodies that can be described by Newtonian mechanics.
Each body has a position ${\bf x}_a$ and fixed mass $m_a$, $a=1,\ldots, N$.
The associated center of mass is    
\begin{equation}\label{Newtbfx}
 {\bf x} = \frac{\sum_a m_a{\bf x}_a}{\sum_a m_a} .
\end{equation}
Now suppose that each body can attain some additional mass $\delta m_a$.
For example, if $m_a$ is the mass of a solid ball, than the ball can gather dust from the environment
and increase its mass by the mass of the dust $\delta m_a$. The total mass of each body (e.g. solid ball plus dust) is then
\begin{equation}
 \tilde{m}_a=m_a+ \delta m_a ,
\end{equation}
and the associated center of mass is
\begin{equation}\label{Newttildex}
 \tilde{\bf x} = \frac{\sum_a \tilde{m}_a{\bf x}_a}{\sum_a \tilde{m}_a} .
\end{equation}

As an extreme example, consider a system of two bodies, for which ${\bf x}_1=0$,
$\tilde{m}_1=m_1\gg \tilde{m}_2$, so that (\ref{Newttildex}) can be approximated with
\begin{equation}\label{Newttildex2}
 \tilde{\bf x} = \frac{\tilde{m}_2{\bf x}_2}{m_1} .
\end{equation}   
Assuming that $m_1$ and ${\bf x}_2$ do not change with time, the velocity of the center of mass is
\begin{equation}
 \frac{d\tilde{\bf x}}{dt}=\frac{1}{m_1} \frac{d\tilde{m}_2}{dt} {\bf x}_2 .
\end{equation}
It has two strange features. First, the center of mass has velocity even though neither of the two body moves.
The velocity of the center of mass is a sole consequence of the change of mass of one of the bodies.
Second, by taking the position $|{\bf x}_2|$ of the second body arbitrarily large, 
the speed of the center of mass can be arbitrarily large as well, in particular it can exceed 
the speed of light. In the next section we shall see that it can exceed the speed of light even 
in a special relativistic context, which is not inconsistent with special relativity 
because no physical body moves faster than light. 

After this extreme example, let us turn back to general analysis. Writing (\ref{Newttildex}) as
\begin{equation}\label{Newttildex3}
 \sum_a \tilde{m}_a \tilde{\bf x} = \sum_a \tilde{m}_a{\bf x}_a 
\end{equation}
and taking the time derivative, we have
\begin{equation}\label{Newttildex4}
 \sum_a \frac{d\tilde{m}_a}{dt} \tilde{\bf x} + \tilde{m} \frac{d\tilde{\bf x}}{dt}
= \sum_a \frac{d\tilde{m}_a}{dt}{\bf x}_a + \sum_a  \tilde{m}_a \frac{d{\bf x}_a}{dt} ,
\end{equation}
where
\begin{equation}
 \tilde{m} = \sum_a \tilde{m}_a .
\end{equation}
Defining 
\begin{equation}\label{Pdefs}
 {\bf P}=\sum_a  \tilde{m}_a \frac{d{\bf x}_a}{dt} , \;\;\;\;
\tilde{\bf P}= \tilde{m} \frac{d\tilde{\bf x}}{dt} ,
\end{equation}
(\ref{Newttildex4}) can be written as
\begin{equation}\label{Newthm}
 {\bf P}=\tilde{\bf P} + \sum_a \frac{d\tilde{m}_a}{dt} (\tilde{\bf x}-{\bf x}_a).
\end{equation}
In (\ref{Pdefs}), ${\bf P}$ is the total momentum of the system, while $\tilde{\bf P}$ is momentum of the center of mass.
When the masses do not change with time, then ${\bf P}$ and $\tilde{\bf P}$ are the same thing, as usual.
From (\ref{Newthm}), however, we see that ${\bf P}$ and $\tilde{\bf P}$ are not the same 
when the masses change with time. They differ by the last term in (\ref{Newthm}),
which we refer to as {\it hidden momentum}
\begin{equation}\label{Phiddef}
 {\bf P}_{\rm hidden}=\sum_a \frac{d\tilde{m}_a}{dt} (\tilde{\bf x}-{\bf x}_a).
\end{equation}

What is the physical momentum, ${\bf P}$ or $\tilde{\bf P}$? From definitions (\ref{Pdefs})
we see that ${\bf P}$ is associated with velocities of the physical bodies $\frac{d{\bf x}_a}{dt}$, 
while $\tilde{\bf P}$ is associated with velocity of the center of mass $\frac{d\tilde{\bf x}}{dt}$.
In the extreme case above, where no physical bodies move, we have ${\bf P}=0$ and  
$\tilde{\bf P}=\frac{d\tilde{m}_2}{dt} {\bf x}_2$. This example suggests that the physical momentum is
${\bf P}$, not $\tilde{\bf P}$.

Another way to see that ${\bf P}$, and not $\tilde{\bf P}$, is physical is as follows. Both ${\bf P}$ and $\tilde{\bf P}$
describe some kind of momentum of the composite body, consisting of $N$ individual bodies.  
Intuitively, the physical momentum is supposed to describe the motion of the composite body as a whole.
But what does the ``body as a whole'' even mean? That notion makes the most sense when the composite body is rigid,
i.e., when there is no relative motion between the individual bodies. In this case all individual bodies have the same velocity 
$\frac{d{\bf x}_a}{dt}$, equal to
\begin{equation}
 \frac{d{\bf x}_a}{dt} = \frac{d{\bf x}}{dt} ,
\end{equation}
where ${\bf x}$ is the center of mass (\ref{Newtbfx}) defined with fixed masses $m_a$ that do not change with time.
This center of mass (\ref{Newtbfx}) thus has the same velocity as the geometrical center $N^{-1}\sum_a {\bf x}_a$ of
the rigid body. On the other hand, the center of mass $\tilde{\bf x}$, defined by (\ref{Newttildex}), 
does not have the same velocity as the geometrical center of the rigid body. 
But the velocity of the rigid body is naturally defined as the velocity of its geometrical center, so we conclude 
that the velocity of the rigid body is $\frac{d{\bf x}}{dt}$, not $\frac{d\tilde{\bf x}}{dt}$.
Thus, if the physical momentum is interpreted as a quantity associated with motion of the composite body as a whole,
then physical momentum is ${\bf P}$, not $\tilde{\bf P}$. In particular, the momentum of the truck at rest 
in the puzzling question from Introduction is the unphysical momentum $\tilde{\bf P}$, not the 
physical momentum ${\bf P}$.

To conclude, the hidden momentum (\ref{Phiddef}) is the quantity that needs to be added to the momentum $\tilde{\bf P}$
of the center of mass, to get the physical momentum ${\bf P}$, as described by (\ref{Newthm}).

\section{Hidden momentum in relativistic mechanics}

In Newtonian mechanics the hidden momentum originated from a possibility to change the mass $\tilde m_a$ of the body, 
by adding matter. In relativistic mechanics there is also an additional way to change the mass, without adding matter.
The so called ``relativistic mass'', often introduced in older literature, as well as in some introductory treatments
of special relativity, depends on the velocity of the body. Denoting the variable ``relativistic mass'' with $\tilde m$,
and the fixed mass of the body at rest with $m$, they are related through the formula
\begin{equation}\label{relm}
 \tilde{m}=\gamma m ,
\end{equation}
where $\gamma=1/\sqrt{1-{\bf v}^2}$, ${\bf v}=\frac{d{\bf x}}{dt}$ is the 3-velocity of the body, 
and we work in units $c=1$.
In the modern covariant language, the mass $m$ is a Lorentz scalar, while the variable mass $\tilde{m}$ is really the energy
$E$, which is the time-component of the 4-momentum $P^{\mu}=(E,{\bf P})$. Thus (\ref{relm}) can be written as 
\begin{equation}\label{relmE}
 E=\gamma m ,
\end{equation}
which is nothing but the time-component of the 4-momentum
\begin{equation}
 P^{\mu}=m\frac{dx^{\mu}}{d\tau} ,
\end{equation}
where $\tau$ is the proper time given by
\begin{eqnarray}
 d\tau^2 &=& dx^{\mu}dx_{\mu} =dx^{0}dx^{0}-dx^{i}dx^{i} = dt^2-d{\bf x}^2
\nonumber \\
&=& dt^2 \left[ 1-\left(\frac{d{\bf x}}{dt}\right)^2 \right] = dt^2 (1-  {\bf v}^2) = \frac{dt^2}{\gamma^2} ,
\end{eqnarray}
and we use the $(+---)$ metric signature.

Now let us consider the system of $N$ relativistic bodies, each having the 4-momentum
\begin{equation}\label{OPmu}
  P_a^{\mu} = m_a \frac{dx_a^{\mu}}{d\tau_a} = m_a \frac{dx_a^{0}}{d\tau_a}  \frac{dx_a^{\mu}}{dx_a^{0}}
 = P_a^0  \frac{dx_a^{\mu}}{dx_a^{0}} = E_a \frac{dx_a^{\mu}}{dt_a} .
\end{equation}
We can define two centers of mass, ${\bf x}$ and $\tilde{\bf x}$, by the same equations
(\ref{Newtbfx}) and (\ref{Newttildex}) as in the Newtonian case.
Eq.~(\ref{Newttildex}) can now be written as
\begin{equation}\label{Newttildexrel}
 \tilde{\bf x} = \frac{\sum_a E_a{\bf x}_a}{\sum_a E_a} ,
\end{equation}
so $\tilde{\bf x}$ in the relativistic case can be interpreted as the center of energy. 
Then, by exactly the same analysis that led to (\ref{Newthm}), we obtain
\begin{equation}\label{Newthmrel}
 {\bf P}=\tilde{\bf P} + \sum_a \frac{dE_a}{dt} (\tilde{\bf x}-{\bf x}_a),
\end{equation}
where
\begin{equation}\label{Pdefsrel}
 {\bf P}=\sum_a  E_a \frac{d{\bf x}_a}{dt} , \;\;\;\;
\tilde{\bf P}= E \frac{d\tilde{\bf x}}{dt} ,
\end{equation}
and $E=\sum_a E_a$ is the total energy. Clearly, ${\bf P}$ is the total 3-momentum, while $\tilde{\bf P}$
is the momentum of the center of energy.

One can argue that the physical momentum is ${\bf P}$, not $\tilde{\bf P}$, by arguments analogous to that in the Newtonian case.
Moreover, in the relativistic case, there is one additional argument based on considering an extreme case analogous to the extreme
Newtonian case. We consider  a system of two bodies, for which ${\bf x}_1=0$,
$E_1=m_1\gg E_2$, so that (\ref{Newttildexrel}) can be approximated with
\begin{equation}\label{Newttildex2rel}
 \tilde{\bf x} = \frac{E_2{\bf x}_2}{m_1} ,
\end{equation}   
which is analogous to (\ref{Newttildex2}).
Hence
\begin{equation}\label{extreme2v}
 \frac{d\tilde{\bf x}}{dt}=\frac{1}{m_1} \frac{dE_2}{dt} {\bf x}_2 + \frac{E_2}{m_1}\frac{d{\bf x}_2}{dt} .
\end{equation}
The velocity of the second body $\frac{d{\bf x}_2}{dt}$ cannot exceed the speed of light. 
Nevertheless, due to the first term on the right-hand side of (\ref{extreme2v}), the velocity of the center of energy
can be arbitrarily large and exceed the speed of light just by taking $|{\bf x}_2|$
arbitrarily large. Thus the momentum $\tilde{\bf P}$ is associated with a velocity of the center of energy 
that can exceed the speed of light, so in general it cannot be associated with a velocity of a physical body. 

\section{Hidden momentum of magnetic (and electric) dipole}

\begin{figure}[!t]
\centering
\begin{tikzpicture}
\draw (0,0) circle (1.5);
\draw (0,0) circle (2);
\draw [semithick, -latex] (0, 1.75) -- (-0.6, 1.75);
\draw [semithick, -latex] (0, -1.75) -- (1.2, -1.75);
\draw [semithick, -latex] (-1.75,0) -- (-1.75, -.9);
\draw [semithick, -latex] (1.75,0) -- (1.75, 1.5);
\draw [fill=lightgray] (90:1.75) circle (0.2);
\draw [fill=lightgray] (00:1.75) circle (0.2);
\draw [fill=lightgray] (180:1.75) circle (0.2);
\node [above] at (-1.7, 0.13) {${\bf x}_a$};
\draw [fill=lightgray] (270:1.75) circle (0.2);
\fill (0,0) circle (0.05) node [above] {$ {\bf x}$};
\fill (.45,-0.3) circle (0.05) node [below] {$\tilde{\bf x}$};
\draw [thick, -latex] (0,0) -- (.45, -0.3) node [midway, above right] {$\delta\tilde{\bf x}$} ;
\draw [thick, -latex] (0,0) -- (-1.75, 0) node [midway, above] {$ {\bf r}_a$};
\draw [thick, -latex] (.45,-0.3) -- (-1.75, 0) node [midway, below] {$\tilde{\bf r}_a$};
\end{tikzpicture}
\caption{Schematic illustration of a current-carrying wire in which the center of mass of charges $\bf x$ disagrees 
with their center of energy $\tilde{\bf x}$ due to a difference in speeds.}
\label{fig1}
\end{figure}
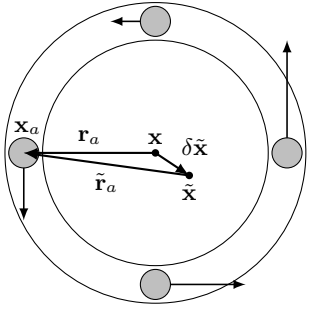

Let us write (\ref{Newthmrel}) as
\begin{equation}\label{Ohidpre}
  {\bf P} = \tilde{\bf P} - \sum_a \frac{dE_a}{dt}\tilde{\bf r}_a ,
\end{equation}
where
$\tilde{\bf r}_a = {\bf x}_a - \tilde{\bf x}$.
We can write
$\tilde{\bf r}_a = {\bf r}_a - \delta \tilde{\bf x}$ ,
where
\begin{equation}\label{radef}
 {\bf r}_a = {\bf x}_a - {\bf x}
\end{equation}
and $\delta \tilde{\bf x} = \tilde{\bf x}-{\bf x}$, 
so (\ref{Ohidpre}) can be written as
\begin{equation}\label{Ohidpre2}
  {\bf P} = \tilde{\bf P} - \sum_a \frac{dE_a}{dt}({\bf r}_a - \delta \tilde{\bf x}) .
\end{equation}
We are considering a magnetic dipole, such as a coil with an electric current, for which $|{\bf r}_a|$ 
is much larger than the difference $|\delta\tilde{\bf x}|$ between the two centers of masses,
as illustrated in Fig.~\ref{fig1}. 
Hence (\ref{Ohidpre2}) can be approximated with   
\begin{equation}\label{Ohid}
  {\bf P} = \tilde{\bf P} - \sum_a \frac{dE_a}{dt}{\bf r}_a .
\end{equation}

To compute $\frac{dE_a}{dt}$, we start from the equations of motion for the particles in the form \cite{jackson}
\begin{equation}\label{eoma}
 m_a\frac{d^2x_a^{\mu}}{d\tau_a^2} = q_aF^{\mu\nu}_a\frac{dx_{a\nu}}{d\tau_a} .
\end{equation}
Here $q_a$ are the particle's charges, $F^{\mu\nu}_a=F^{\mu\nu}(x_a)$, and
$F^{\mu\nu}=\partial^{\mu}A^{\nu}-\partial^{\nu}A^{\mu}$ is a fixed background electromagnetic field tensor.
In reality there are also internal forces on the charges that keep different parts of the dipole together,
but we do not consider them explicitly because we assume that all internal forces are in equilibrium, 
so do not contribute to acceleration of the charges.  
Using (\ref{OPmu}) and taking $\mu=0$, (\ref{eoma}) becomes
\begin{eqnarray}\label{dEadtau}
\frac{dE_a}{d\tau_a} &=& q_aF^{0i}_a\frac{dx_{ai}}{d\tau_a}
\nonumber \\
&=& q_a E^i_a \frac{dx_{a}^i}{d\tau_a} ,
\end{eqnarray}
where in the second line we have used $F^{0i}_a=-E^i_a$ (with $E^i$ being the components of the electric field)
and $\frac{dx_{ai}}{d\tau_a}=-\frac{dx_{a}^i}{d\tau_a}$.
Using also
\begin{equation}\label{tauat}
 \frac{d}{d\tau_a}= \frac{dt}{d\tau_a} \frac{d}{dt}=\gamma_a \frac{d}{dt} ,
\end{equation}
(\ref{dEadtau}) can be written as 
\begin{equation}\label{OEt}
\frac{dE_a}{dt}= q_aE^{i}_a\frac{dx_{a}^i}{dt} ,
\end{equation}
so (\ref{Ohid}) in the component form reduces to 
\begin{equation}\label{Ohid2}
P^j =  \tilde{P^j} - \sum_a q_aE^{i}_a\frac{dx_{a}^i}{dt} r_a^j .
\end{equation}
 
Now we perform the multipole expansion in the second term of (\ref{Ohid2})
and keep only the zeroth and first term in the expansion. 
But the multipole expansion is really the expansion in $r_a^j$, so $E^{i}_a$ should be expanded only to the zeroth order,
i.e., $E^{i}_a\equiv E^{i}(x_a)$ should be approximated with  $E^{i}(x)\equiv E^i$. From (\ref{radef}) we see that 
\begin{equation}\label{velocity}
 \frac{dx_{a}^i}{dt} = \frac{dx^i}{dt} + \frac{dr_{a}^i}{dt} ,
\end{equation}
so (\ref{Ohid2}) becomes
\begin{equation}\label{Ohid3}
P^j =  \tilde{P^j} - E^{i}\frac{dx^i}{dt} \sum_a q_a r_a^j - E^{i} \sum_a q_a r_a^j \frac{dr_a^i}{dt} .
\end{equation}
For simplicity we assume that the electric dipole moment is zero $\sum_a q_a r_a^j=0$, 
while in the last term 
only the antisymmetric part contributes, 
as we have shown in Appendix, Eqs.~(\ref{sym+asym})-(\ref{sym0}). 
Hence (\ref{Ohid3}) reduces to
\begin{equation}\label{Ohid4}
P^j =  \tilde{P^j} + E^{i} \sum_a q_a \frac{1}{2}\left( r_a^i \frac{dr_a^j}{dt} - r_a^j \frac{dr_a^i}{dt} \right),
\end{equation}
or 
\begin{equation}\label{totmomdipole}
 {\bf P}= \tilde{\bf P} + {\bf P}_{\rm hidden} ,
\end{equation}
where
\begin{equation}\label{dipolehidmom}
{\bf P}_{\rm hidden} =
\sum_a q_a \frac{1}{2} 
[ ({\bf E}\cdot{\bf r}_a) \dot{\bf r}_a - {\bf r}_a({\bf E}\cdot \dot{\bf r}_a)] ,
\end{equation}
and $\dot{\bf r}_a=\frac{d{\bf r}_a}{dt}$. Using the 3-vector identity \cite{jackson}
\begin{equation}
({\bf E}\cdot{\bf r}_a) \dot{\bf r}_a - {\bf r}_a({\bf E}\cdot \dot{\bf r}_a)
= ({\bf r}_a \times \dot{\bf r}_a) \times {\bf E} ,
\end{equation}
(\ref{dipolehidmom}) can be written as
\begin{equation}\label{dipolehidmomfin}
 {\bf P}_{\rm hidden} = \mbox{\boldmath$\mu$} \times {\bf E} ,
\end{equation}
where $\mbox{\boldmath$\mu$}$ is the magnetic dipole moment 3-vector defined in (\ref{defmu}).

Formula (\ref{dipolehidmomfin}) is known in the literature \cite{hnizdo1992, HausPenfield1968, vaidman} as the hidden momentum
of the magnetic dipole.
From (\ref{ap10}), (\ref{totmomdipole}) and (\ref{dipolehidmomfin}) we see that 
\begin{equation}\label{NeomPti3}
   \frac{d\tilde{\bf P}}{dt}= \mbox{\boldmath$\nabla$} (\mbox{\boldmath$\mu$} \cdot {\bf B})
 - \frac{d}{dt}(    \mbox{\boldmath$\mu$}     \times {\bf E} ).
\end{equation}
In the literature it is sometimes interpreted 
\cite{hnizdo1992,vaidman} as a claim that the full ``physical'' force (``physical'' in the sense that it purportedly describes 
the motion of the center of mass) on the magnetic dipole is
$\mbox{\boldmath$\nabla$} (\mbox{\boldmath$\mu$} \cdot {\bf B}) - \frac{d}{dt}( \mbox{\boldmath$\mu$}  \times {\bf E} )$. 
This would make sense
if $\tilde{\bf P}$ could be interpreted as the ``physical'' momentum. 
However, we have argued that $\tilde{\bf P}$ is not the physical momentum. 
The physical momentum is ${\bf P}$, so the physical force is just 
$\mbox{\boldmath$\nabla$} (\mbox{\boldmath$\mu$} \cdot {\bf B})$.

Finally note that (\ref{Ohid3}) can also be evaluated when the electric dipole moment is not zero. 
In this case the second term in (\ref{Ohid3}) is
\begin{equation}
 - E^{i}\frac{dx^i}{dt} \sum_a q_a r_a^j = -({\bf E}\cdot{\bf v}) \mbox{\boldmath$\pi$}^j ,
\end{equation}
where
\begin{equation}
 \mbox{\boldmath$\pi$}=\sum_a q_a {\bf r}_a
\end{equation}
is the electric dipole moment. Hence the full hidden momentum is an extension of (\ref{dipolehidmomfin})
\begin{equation}\label{dipolehidmomfin2}
 {\bf P}_{\rm hidden} =  \mbox{\boldmath$\mu$} \times {\bf E} -\mbox{\boldmath$\pi$}({\bf v}\cdot {\bf E}) .
\end{equation}
The physical momentum in the presence of both magnetic and electric dipole moment
obeys the equation of motion (\ref{ap11gen}).

\section{Conclusion}

We have shown that hidden momentum is not some obscure artifact of the mathematical description but a genuine physical quantity that can arise in both Newtonian and relativistic systems. In non-relativistic contexts, a changing mass distribution can create a discrepancy between the momentum of the center of mass and the total physical momentum, revealing momentum that is ‘‘hidden’’ in the internal dynamics. Under special relativity, this idea naturally extends to the difference between the total 3-momentum and the momentum of the center of energy.

A key illustration is provided by magnetic dipoles, where the hidden momentum appears in the interplay between electromagnetic fields and the internal currents or charge distributions responsible for the dipole moment. Although the center of energy might move in a way that suggests superluminal motion, no actual material body exceeds the speed of light, thus preserving consistency with fundamental physical principles.

By clarifying the origins and manifestations of hidden momentum, we have demystified its role in systems ranging from 
simple mechanical analogies to intricate electromagnetic configurations. 
We hope that this treatment of hidden momentum will serve as a clear guide to its broader implications and help 
avoid common misconceptions in future studies.

\section*{Author Declarations}

The authors have no conflicts to disclose.

\appendix

\section{Equation of motion for the total momentum}

The equations of motion (\ref{eoma}), together with the definition of 4-momenta (\ref{OPmu}), are 
\begin{equation}\label{eoma2}
 \frac{dP_a^{\mu}}{d\tau_a} = q_aF^{\mu\nu}_a\frac{dx_{a\nu}}{d\tau_a}. 
\end{equation}
Using (\ref{tauat}), this becomes 
\begin{equation}\label{eoma3}
 \frac{dP_a^{\mu}}{dt} = q_aF^{\mu\nu}_a\frac{dx_{a\nu}}{dt}.
\end{equation}
The time component of this is (\ref{OEt}). Here we are interested in the spatial part $\mu=i$ summed over all $a$, so
\begin{equation}\label{ap3}
 \frac{dP^{i}}{dt} = \sum_a q_aF^{i\nu}_a\frac{dx_{a\nu}}{dt} , 
\end{equation}
where $P^{i}=\sum_a P_a^{i}$ is the total momentum. 
Next we write (\ref{ap3}) as
\begin{equation}
 \frac{dP^{i}}{dt} = \sum_a q_aF^{i0}_a\frac{dx_{a0}}{dt} + \sum_a q_aF^{ij}_a\frac{dx_{aj}}{dt}, 
\end{equation}
so using $F^{i0}_a=-E^i_a$, $\frac{dx_{a0}}{dt}=1$, $F^{ij}_a=-\epsilon^{ijk}B^k_a$ (with $\epsilon^{123}=1$), and
$\frac{dx_{aj}}{dt}=-\frac{dx_{a}^j}{dt}$, we have
\begin{equation}\label{ap5}
 \frac{dP^{i}}{dt} = \sum_a q_a E^{i}_a  + \sum_a q_a \epsilon^{ijk}B^k_a \frac{dx_{a}^j}{dt}, 
\end{equation}
or in the more familiar form
\begin{equation}
 \frac{d{\bf P}}{dt} = \sum_a q_a ({\bf E}_a + {\bf v}_a\times {\bf B}_a) .
\end{equation}

Any function $f_a\equiv f(x_a)$ can be expanded as
\begin{equation}
 f(x_a)=f(x)+(x_a^l-x^l)\partial_l f(x) +\ldots ,
\end{equation}
or more compactly $f_a=f + r_a^l f_{,l}$, where we neglect the higher terms in the expansion. 
Thus
\begin{equation}
 E^{i}_a=E^{i} +r_a^l E^{i}_{,l}, \;\;\; B^{k}_a=B^{k} +r_a^l B^{k}_{,l},
\end{equation}
which we insert into (\ref{ap5}), and use (\ref{velocity}) in the form
\begin{equation}
\frac{dx_{a}^j}{dt} = \frac{dx^j}{dt} +\frac{dr_{a}^j}{dt} ,
\end{equation}
to obtain 
\begin{eqnarray}\label{ap6}
 \frac{dP^{i}}{dt} &=& E^i\sum_a q_a  + E^{i}_{,l}\sum_a q_a r_a^l
 \\
& &   + \epsilon^{ijk}B^k \frac{dx^j}{dt} \sum_a q_a   + \epsilon^{ijk}B^k \sum_a q_a \frac{dr_{a}^j}{dt} 
 \nonumber \\
& & + \epsilon^{ijk}B^{k}_{,l} \frac{dx^j}{dt} \sum_a q_a r_a^l 
  +\epsilon^{ijk}B^{k}_{,l}  \sum_a q_a r_a^l \frac{dr_{a}^j}{dt} .
 \nonumber
\end{eqnarray}
This looks rather cumbersome, but it simplifies a lot 
if we assume that the total charge and total electric dipole moment vanish
\begin{equation}\label{zeroqpi}
 \sum_a q_a =0, \;\;\; \sum_a q_a r_a^l =0,
\end{equation}
because then only the last term survives
\begin{equation}\label{ap7}
 \frac{dP^{i}}{dt} = \epsilon^{ijk}B^{k}_{,l}  \sum_a q_a r_a^l \dot{r}_{a}^j ,
\end{equation}
where the dot denotes the derivative over time $t$.

Next note that 
\begin{equation}\label{sym+asym}
 r_a^l \dot{r}_{a}^j =\frac{1}{2} (r_a^l \dot{r}_{a}^j + r_a^j \dot{r}_{a}^l)
+ \frac{1}{2} (r_a^l \dot{r}_{a}^j - r_a^j \dot{r}_{a}^l) ,
\end{equation}
and that the symmetric part is a total derivative
\begin{equation}
\frac{1}{2} (r_a^l \dot{r}_{a}^j + r_a^j \dot{r}_{a}^l) = \frac{1}{4} \frac{d}{dt}(r_a^l r_{a}^j + r_a^j r_{a}^l) .
\end{equation}
We assume that in the proximity of any charge $q_a$ at the position ${\bf r}_a$ there is an opposite charge
$q_{a'}=-q_a$ at nearly the same position ${\bf r}_{a'}\simeq {\bf r}_a$, so that 
\begin{equation}\label{no-quadrupole}
 \sum_a q_a (r_a^l r_{a}^j + r_a^j r_{a}^l) \simeq 0 .
\end{equation}
Hence
\begin{equation}\label{sym0}
\sum_a q_a \frac{1}{2} (r_a^l \dot{r}_{a}^j + r_a^j \dot{r}_{a}^l) = 
\frac{1}{4} \frac{d}{dt} \sum_a q_a (r_a^l r_{a}^j + r_a^j r_{a}^l) \simeq 0
\end{equation}
is negligible, so
only the antisymmetric part of (\ref{sym+asym}) contributes to (\ref{ap7})
\begin{equation}\label{ap8}
 \frac{dP^{i}}{dt} = \epsilon^{ijk}B^{k}_{,l}  \frac{1}{2}\sum_a q_a (r_a^l \dot{r}_{a}^j-r_a^j \dot{r}_{a}^l) .
\end{equation}

Next we use the identity
\begin{equation}
 r_a^l \dot{r}_{a}^j-r_a^j \dot{r}_{a}^l = \epsilon^{ljm}({\bf r}_a\times\dot{\bf r}_a)^m ,
\end{equation}
the definition of magnetic dipole moment
\begin{equation}\label{defmu}
 \mbox{\boldmath$\mu$}=\frac{1}{2}\sum_a q_a{\bf r}_a\times \dot{\bf r}_a ,
\end{equation}
and the identity
\begin{equation}
 \epsilon^{ijk}\epsilon^{ljm}=\delta^{il}\delta^{km}-\delta^{im}\delta^{kl} ,
\end{equation}
to write (\ref{ap8}) as 
\begin{equation}\label{ap9}
 \frac{dP^{i}}{dt} = \epsilon^{ijk}\epsilon^{ljm} B^{k}_{,l} \mu^m
=B^{k}_{,i}\mu^k-B^{k}_{,k}\mu^i .
\end{equation}
But $B^{k}_{,k}= \mbox{\boldmath$\nabla$}\cdot{\bf B}=0$, and $\partial_i\mu^k=0$ because the particle positions
${\bf r}_a(t)$ in (\ref{defmu}) do not have an explicit dependence on $x^i$, so (\ref{ap9}) can be written in the final form
\begin{equation}\label{ap10}
  \frac{d{\bf P}}{dt} = \mbox{\boldmath$\nabla$} (\mbox{\boldmath$\mu$} \cdot {\bf B}) .
\end{equation}
In the literature, (\ref{ap10}) is usually derived for continuous distributions of charge \cite{jackson,boyer},
while here we have derived it for point charges.

Finally, for completeness, let us evaluate all the other terms in (\ref{ap6}) without assuming (\ref{zeroqpi}).
Instead of (\ref{zeroqpi}), we define the total charge and total electric dipole moment as
\begin{equation}\label{qpi}
 q=\sum_a q_a, \;\;\; \pi^l=\sum_a q_a r_a^l .
\end{equation}
Then all the other terms in (\ref{ap6}) are
\begin{equation}
 E^i\sum_a q_a =qE^i,
\end{equation}
\begin{equation}
 E^{i}_{,l}\sum_a q_a r_a^l = \pi^l E^{i}_{,l} = (\mbox{\boldmath$\pi$}\cdot\mbox{\boldmath$\nabla$})E^{i},
\end{equation}
\begin{equation}
 \epsilon^{ijk}B^k \frac{dx^j}{dt} \sum_a q_a = q \epsilon^{ijk} v^j B^k = q ({\bf v}\times{\bf B})^i ,
\end{equation}
\begin{equation}
\epsilon^{ijk}B^k \sum_a q_a \frac{dr_{a}^j}{dt} = \epsilon^{ijk} \dot{\pi}^j B^k 
= (\dot{\mbox{\boldmath$\pi$}} \times {\bf B})^i ,
\end{equation}
\begin{eqnarray}
  \epsilon^{ijk}B^{k}_{,l} \frac{dx^j}{dt} \sum_a q_a r_a^l 
&=& \pi^l \epsilon^{ijk} v^j B^{k}_{,l} = \pi^l\partial_l (\epsilon^{ijk}v^j B^{k})
\nonumber \\
&=& (\mbox{\boldmath$\pi$}\cdot\mbox{\boldmath$\nabla$}) ({\bf v}\times{\bf B})^i .
\end{eqnarray}
Thus, collecting all the terms in (\ref{ap6}) together,  
(\ref{ap10}) generalizes to
\begin{equation}\label{ap11gen}
  \frac{d{\bf P}}{dt} = [q+\mbox{\boldmath$\pi$}\cdot\mbox{\boldmath$\nabla$}] ({\bf E} + {\bf v}\times{\bf B})
+ \mbox{\boldmath$\nabla$} (\mbox{\boldmath$\mu$} \cdot {\bf B}) + \dot{\mbox{\boldmath$\pi$}} \times {\bf B}.
\end{equation}
But actually, since we neglected (\ref{no-quadrupole}) by assuming that in the proximity of any charge there is also 
an equal amount of opposite charge, 
by the same assumption it follows that the electric dipole moment should also be 
neglected, $\mbox{\boldmath$\pi$} \simeq 0$. With that assumption, (\ref{ap11gen}) simplifies to
\begin{equation}\label{ap11}
  \frac{d{\bf P}}{dt} = q ({\bf E} + {\bf v}\times{\bf B})
+ \mbox{\boldmath$\nabla$} (\mbox{\boldmath$\mu$} \cdot {\bf B}) .
\end{equation}

\end{document}